\documentclass[runningheads]{llncs}
\usepackage[T1]{fontenc}
\usepackage{graphicx}
\usepackage{cite}
\usepackage{amsmath,amssymb,amsfonts}
\usepackage{algorithmic}
\usepackage{graphicx}
\usepackage{url}
\usepackage{textcomp}
\usepackage{xcolor}
\usepackage{todonotes}
\usepackage{pifont}
\usepackage{comment}
\usepackage{orcidlink}
\usepackage{booktabs}       % for table
\usepackage{multicol}       % for table
\usepackage{multirow}       % for table
\usepackage{subcaption}
\def\BibTeX{{\rm B\kern-.05em{\sc i\kern-.025em b}\kern-.08em
    T\kern-.1667em\lower.7ex\hbox{E}\kern-.125emX}}
\begin{document}
\title{Energy-Aware Quantum-Enhanced Computing Continuum}
%  Changement of title form the original title A Carbon-Aware Framework for AI Workloads Migration Across the Computing Continuum

\author{
Carlos J. BARRIOS H.\inst{1,2,3}\orcidID{0000-0002-3227-8651} \\ \and
Frédéric LE MOUËL\inst{3}\orcidID{0000-0002-7323-4057}  \\ \and
Oscar CARRILLO\inst{4}\orcidID{0000-0001-5081-1774}
}
\authorrunning{Barrios H. et al.}
\institute{UIS, SC3UIS, CAGE, Bucaramanga, Colombia \and 
LIG/INRIA-Grenoble, Saint Martin d'Hères, France \and 
INSA Lyon, INRIA, CITI, UR3720, Villeurbanne, France 
\and CPE, INSA Lyon, INRIA, CITI, UR3720, Villeurbanne, France \
}
%\institute{Princeton University, Princeton NJ 08544, USA \and
%Springer Heidelberg, Tiergartenstr. 17, 69121 Heidelberg, Germany
\maketitle              % typeset the header of the contribution
\begin{abstract}
We discuss a Quantum-Enhanced Computing Continuum, a heterogeneous, hybrid architecture that integrates quantum processing units (QPUs) within an Edge-Cloud-HPC fabric. Promote sustainability by shifting from performance to "energy-aware integration.' The architecture has three layers: a Physical Layer with shared fiber-optic infrastructure, a Control and Orchestration Layer managed by the user, and an Application Layer with an  Adaptive Quantum Classical Fusion (AQCF) framework. Tighter system integration, like moving from cloud coupling to cryogenic logic, reduces energy waste and "thermal footprints.' The aim is a Green Performance Advantage: energy per problem solved in the era of Advanced Computing.
\end{abstract}
\section{Introduction}

The integration of Quantum Computing into the Computing Continuum involves using quantum devices as specialized co-processors or accelerators, with the Quantum Processor Unit (QPU) analogous to other accelerators, such as graphics processing units (GPUs). This move was driven more by the need to support modern distributed applications, such as artificial intelligence (AI), Big Data, and scientific computing, than by technological advances alone. It also requires integrating quantum computing devices and emerging non-Von Neumann architectures. Nonetheless, as noted by various authors, there are significant challenges across different architectural layers and with respect to sustainability \cite{b1}.

This merge of Edge, Fog, Cloud, and High-Performance Computing (HPC) into a single framework includes mobile devices and distributed Quantum Processing Units (QPUs) within the edge-cloud continuum. This evolution extends quantum computing beyond simple cloud access by introducing a distributed inference engine that employs hybrid classical-quantum neural networks (QNNs) and employs techniques such as circuit cutting and classical partitioning. These methods extend classical split-computing approaches to more effectively solve problems across heterogeneous resources. This transition shifts from theoretical concepts to real-time, mission-critical applications where latency and data sovereignty are crucial. For example, in Autonomous Vehicle Swarms, localized QPUs handle path planning and collision avoidance at the edge, bypassing cloud delays \cite{b1a}. In Smart Grid Management, quantum sensors accurately track fluctuations, enabling rapid load balancing to prevent failures \cite{b1b}. In Privacy-Preserving Healthcare \cite{b1c} \cite{b1d}, circuit cutting allows devices to process sensitive data locally, with only intensive computations sent to the quantum cloud to safeguard privacy. Incorporating quantum intelligence into Industrial IoT enables real-time stress analysis and anomaly detection at manufacturing sites, making quantum technology a widespread utility for physical systems \cite{b1e}. Similarly, the integration of artificial intelligence and quantum computing across various applications has emerged as a pivotal development in the progression of Industry 6.0. These technologies are propelling advancements in automation, advanced analytics, and process optimization. Their combination has the potential to transform sectors such as data science, healthcare, finance, and cybersecurity by enabling faster, more efficient computations through qubits, superposition, and quantum entanglement \cite{b1f}. These advances facilitate the use of key quantum devices within this continuum, merging quantum cloud, edge, and sensing technologies into a new architectural paradigm, as shown in Table \ref{tab:QEC}.

\begin{table}[htbp!]
    \centering
        \renewcommand{\arraystretch}{1.5} % Adds padding between rows for readability

    \caption{Quantum at Edge and Cloud}
    \label{tab:QEC}
    \begin{tabular}{|p{1.5cm}|p{4.0cm}|p{6.2cm}|}
        \hline
        \scriptsize{\textbf{Tier}} & \scriptsize{\textbf{Device Type}} & \scriptsize{\textbf{Tight Integration (Use Case)}} \\ 
        \hline
        \scriptsize{Quantum Cloud} & \scriptsize{Large-Scale, Cryongenic QPUs (1000+ Qubits)} & \scriptsize{Molecular Simulation, Deep-tier Financial Optimization, QML (Training/Inference)} \\
        \hline
        \scriptsize{Quantum Edge}  & \scriptsize{Mobile/Compact QPUs} & \scriptsize{Real Time Signal Processing, Secure QKD for IoT, QML (Federated Inference)} \\
        \hline
        \scriptsize{Quantum Sensing}   & \scriptsize{High Precision Sensors} & \scriptsize{Sub-Surface mapping, biomedical, imaging in the Field} \\ 
        \hline
        
    \end{tabular}
\end{table}

\begin{figure}[!htbp]
  \begin{center}
    \includegraphics[width=0.58\textwidth]{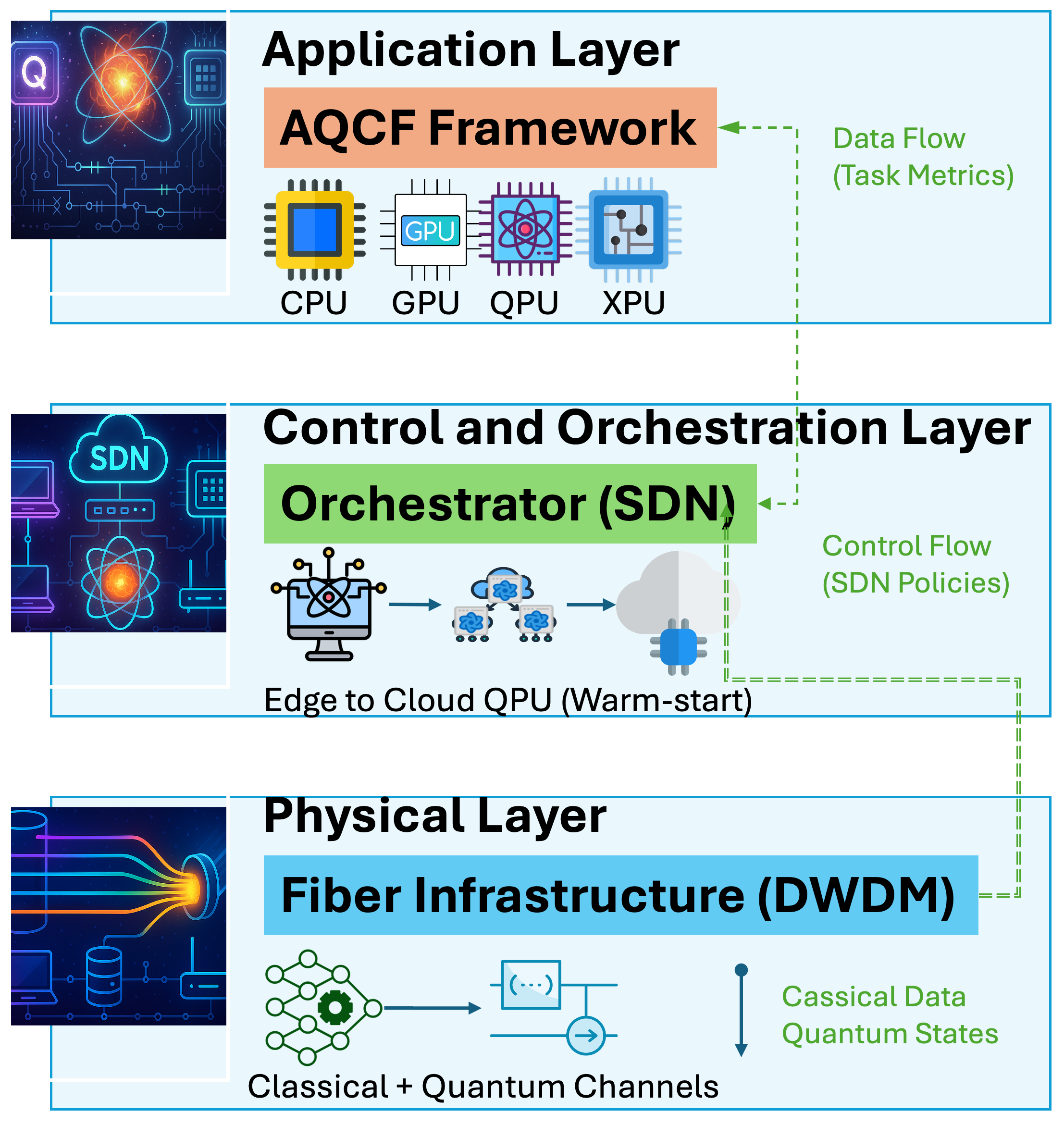}
  \end{center}
  \caption{Quantum Classic Hybrid Architecture}
  \label{fig:QCHA}
\end{figure}

Table \ref{tab:QEC} presents the tier, common device type, and use case. Depending on the use case, specific devices are engaged, such as Quantum Processing Units (QPU), Quantum Networking Units (QNUs), Microwave-to-Optical Transducers, or Quantum Memories. When classical computing intersects with quantum computing, various models and architectures are introduced to address particular issues. The focus is on their role in enhancing data processing efficiency and supporting the industrialization of quantum algorithms by integrating quantum and classical computational models \cite{b1g}. Similarly, it is feasible to position the device within a quantum-classical hybrid architecture, as illustrated in Figure \ref{fig:QCHA}.

The quantum-classical hybrid architecture in Figure \ref{fig:QCHA} has three layers with clear control and data flows. The Physical Layer uses fiber-optic cables with Dense Wavelength Division Multiplexing (DWDM), to separate quantum and classical signals. DWDM is a key technology for enabling high-capacity and flexible quantum communication networks. In addition, to realize the emerging quantum internet, quantum frequency conversion is also essential for bridging different quantum systems over optical fiber networks. Classical channels handle communication and management, while quantum channels carry quantum states on separate wavelengths, even over shared infrastructure. The Control and Orchestration Layer manages resources, using, for example, Software-Defined Networking (SDN), with arrows indicating control signals, network paths, resource allocation, and data exchanges, such as metrics and feedback. When edge devices detect workload-intensive tasks, the orchestrator sends a warm-start command to activate a cloud QPU, reducing latency. The Application Layer's AQCF system evaluates tasks and dynamically splits algorithms across CPUs, GPUs, or QPUs based on real-time feedback, enabling efficient hybrid execution while abstracting infrastructure complexity.

Analyzing Figure \ref{fig:QCHA} raises a key question: how to promote energy-aware integration across application, control, and physical layers to improve performance sustainably, measured by energy per problem solved rather than raw capacity? This proposal examines each layer of a Distributed Quantum Computing framework and integrates Quantum Computing components into the Computing Continuum. Section \ref{sec:BRW} reviews related research on architecture and integration of Quantum and Classical Computing, focusing on middleware and energy-based performance. Section \ref{sec:DQC} describes the Distributed Quantum Computing Framework abstraction. Section \ref{sec:EEGA} discusses Energy Efficiency and Green benefits in Quantum-Classical Hybrid Architectures. Finally, Section \ref{sec:FWC} presents further work and conclusions.

\section{Background and Related Work}
\label{sec:BRW}

In the emergence of federated architectures, for example, in machine learning, the integration of the advantages of classical computing with the power of quantum technologies enables the handling of high-dimensional, complex data, for example, using Quantum Federated Learning \cite{b3} \cite{b4}.  

Visualize the computing continuum as an unbroken chain linking data collection at the 'Edge' to the high-capacity 'Cloud.' To manage this complexity, federated architectures and federated programming models provide a unified way to coordinate diverse resources, enabling dynamic allocation and seamless communication across domains \cite{b5} \cite{b5b} \cite{b6}. This approach simplifies development while enhancing scalability, responsiveness, and resilience. Instead of sending every raw data point to a distant central server, which can be slow, costly, and pose privacy risks, a federated architecture enables local processing and intelligence \cite{b7} \cite{b8}. Specifically, in AI applications, this setup allows devices to collaborate on complex tasks or train AI models while keeping sensitive data local. Such a system creates a smarter, faster, and more secure core system for any digital environment, ensuring computational resources are always available exactly where needed.

Conversely, the primary concerns regarding quantum-classical integration, particularly with regard to the emerging requirements of specific workloads in applications, such as machine learning systems that utilize Quantum Machine Learning (QML) as a system architecture opportunity, pertain not solely to the speed of implementation but also to the sustainability and environmental impact of this hybrid quantum-classical continuum, posing important challenges in implementation that demand a synergistic approach to solving these \cite{b9}. Energy concerns are well known, and there are different models that can already be applied to any infrastructure, including quantum computers and large-scale systems \cite{b10}. The emphasis on energy and carbon footprint factors stems from the fact that, although quantum computers demand substantial energy for cooling, their operational energy consumption or operational energy efficiency for specific problems is significantly lower, by orders of magnitude, compared to classical HPC systems \cite{b11}.

\begin{figure}[!htbp]
  \begin{center}
    \includegraphics[width=0.60\textwidth]{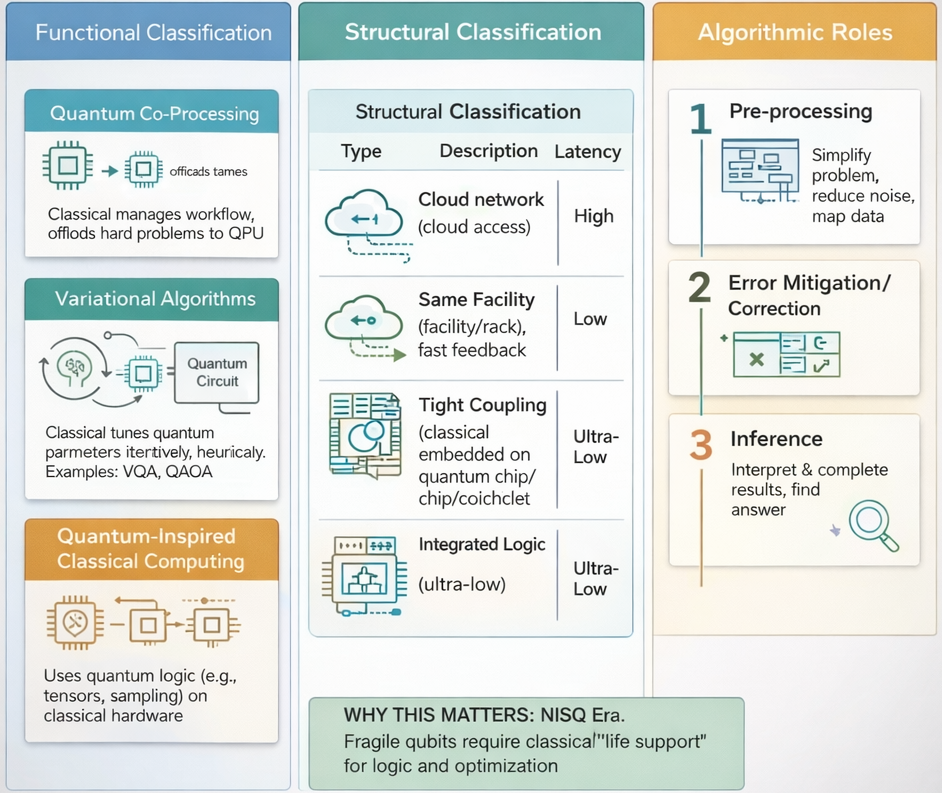}
  \end{center}
  \caption{Hybrid Quantum-Classical Computing Classification}
  \label{fig:HQCPD}
\end{figure}

Depending on the dimensions to observe, it is possible to propose a classification of the results of integrating the components by considering the functional classification, the structural classification (tightness of integration, defining the physical and data-link proximity between the quantum and classical systems), and the algorithmic roles \cite{b12}, as shown in Figure \ref{fig:HQCPD}.

Figure \ref{fig:HQCPD} categorizes hybrid quantum-classical computing into three main dimensions: Function, Structure, and Role. The Function column details workload division, from basic co-processing to the iterative feedback loops in variational algorithms such as VQE. The Structure section defines the physical connections, ranging from high-latency 'Loose Coupling' over the cloud to 'Integrated Logic,' where classical controllers are embedded within the quantum hardware to minimize latency. The Algorithmic Roles segment outlines the workflow: classical computers handle data preparation (pre-processing), qubit error management (mitigation), and the interpretation of probabilistic outcomes (post-processing). Together, these classifications show that classical systems are not merely auxiliary; they serve as essential 'life support' and optimization mechanisms for modern quantum processors.

Quantum-Classical Computing Hybrid Systems (QCCHS) exhibit unique energy patterns, with efficiency depending on interface control. Transitioning from cloud to integrated cryogenic logic reduces energy waste by eliminating cooling "idle power" during network delays. Physical proximity enhances sustainability, as faster feedback loops enable shorter quantum processor operation, reducing the power required for qubits. Classical pre-processing streamlines energy consumption by simplifying the problem, yielding fewer and more efficient quantum operations. Improved integration reduces the "thermal footprint," bringing these systems closer to achieving true energy benefits relative to HPC systems.

QCCHS exhibit distinct energy patterns, with efficiency that depends on interface management. In the Structural dimension, moving from cloud links to cryogenic logic reduces energy waste by eliminating 'idle power' during cooling caused by network delays. Closer physical integration improves sustainability by enabling faster feedback in variational algorithms and lowering power needed to maintain qubits. Classical pre-processing reduces energy consumption by refining the problem formulation, thereby enabling quantum hardware to execute fewer, more efficient operations. Overall, better integration minimizes thermal footprint and provides energy advantages over supercomputers. Understanding application scheduling is crucial, as it influences orchestration and middleware, like in classical computing.  These traits, naturally, affect how these systems are utilized. To fully unlock the environmental advantages of quantum-enhanced computing continuity systems, users must possess the knowledge to address the related challenges\cite{b15}.

\section{Distributed Quantum Computing}
\label{sec:DQC}

Distributed Quantum Computing (DQC) shifts focus from single, large processors to interconnected networks of \emph{systems of systems} \cite{b16}. The primary challenge is establishing remote entanglement between physically separated Quantum Processing Units (QPUs). Using quantum repeaters and microwave-to-optical transducers, DQC links small, modular quantum nodes via fiber-optic connections, thereby combining their qubits to perform calculations that surpass those of individual devices. This system relies heavily on Quantum Teleportation for data transfer and distributed gate operations, reducing decoherence risks over long distances.

Managing this DQC framework can be achieved using Software-Defined Networking (SDN) controllers, which treat the network as a unified computational fabric, handling control signals, network paths, and resource allocation \cite{b17}. However, other management layers are also viable, each with its own pros and cons. In our approach, deploying SDN as an enabler, the Interaction with the Adaptive Quantum Classical Fusion (AQCF) framework allows for high-level coordination by dynamically distributing algorithms across CPUs, GPUs, or QPUs based on real-time task metrics. Federated programming models and architectures facilitate the coordination of diverse resources across different domains, ensuring smooth communication and flexible resource allocation while preserving data sovereignty. Collectively, these systems in a quantum hybrid computing model support scalable, fault-tolerant execution within a heterogeneous grid that promotes sustainability through a \emph{Green} initiative.

\begin{figure}[!htbp]
  \begin{center}
    \includegraphics[width=0.60\textwidth]{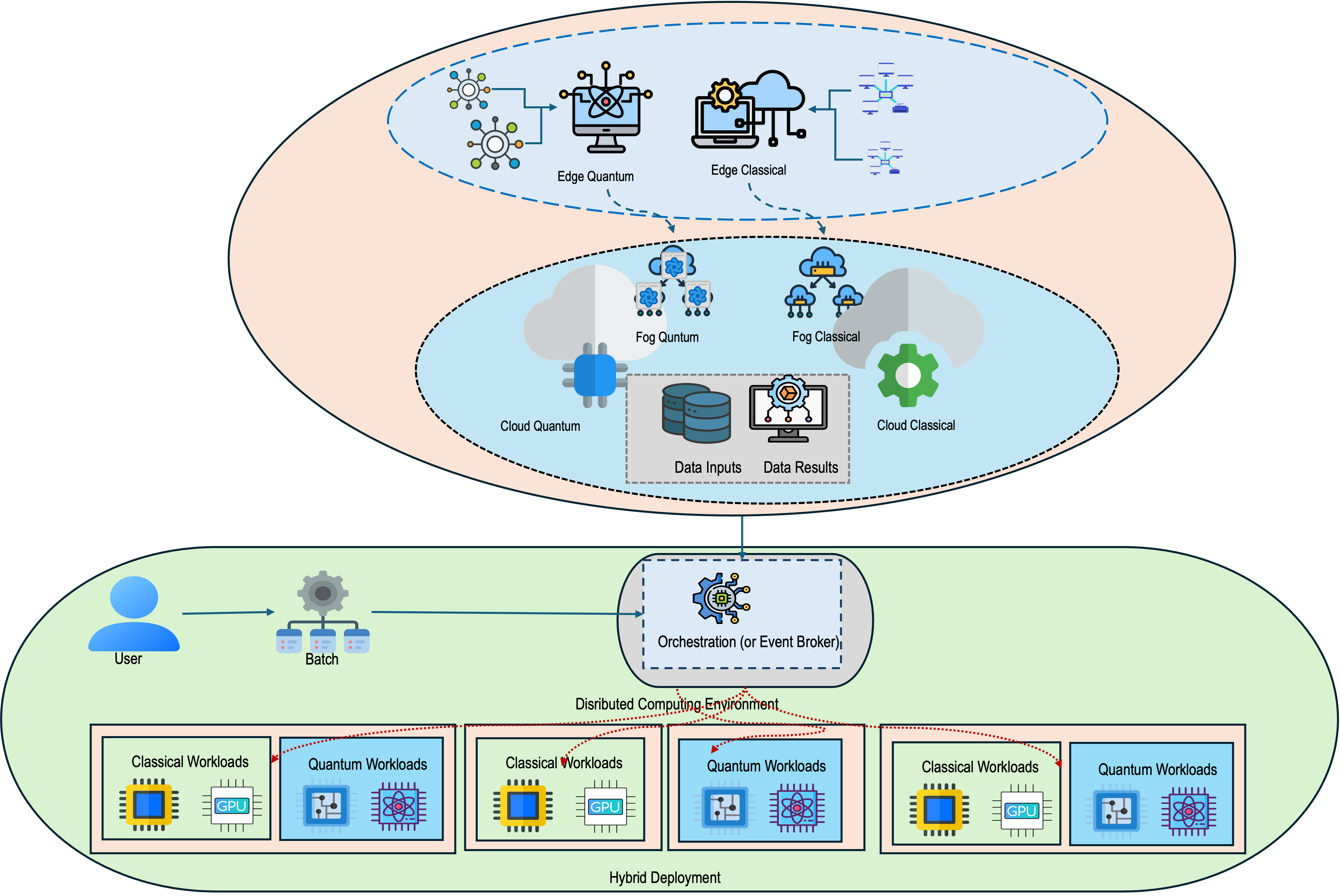}
  \end{center}
  \caption{Distributed Quantum Computing Framework Abstraction}
  \label{fig:DCQHP}
\end{figure}

Observing quantum edge topology, compact or mobile Quantum Processing Units (QPUs) are strategically positioned near data sources to facilitate real-time processing and reduce communication latency. They frequently employ a Tight Coupling configuration, wherein quantum and classical components are co-located to enable rapid feedback. For instance, in an Autonomous Vehicle Swarm, localized QPUs handle path planning and collision avoidance, thereby bypassing cloud latency (as shown in \cite{b1a}). These edge nodes interface with classical hardware via an Adaptive Quantum Classical Fusion (AQCF) system that dynamically allocates workloads, permitting local processing of sensitive information through circuit cutting and offloading computationally intensive tasks to a large cryogenic QPU hosted in the cloud. Figure \ref{fig:DCQHP} depicts a Distributed Quantum Computing Framework that organizes processing infrastructure in a hierarchical structure. This hierarchy begins at the Edge (local devices), progresses through the Fog (intermediate nodes), and culminates at the Cloud (centralized clusters), with each tier equipped with both classical and quantum capabilities. Data traverses these levels and connects to a central Cloud hub responsible for Data Inputs and Data Results.

The central Orchestration (or Event Broker) functions as the system's control center, handling Batch requests from users and distributing them within a Distributed Computing Environment. Figure \ref{fig:DCQHP}. In the bottom Hybrid Deployment layer, workloads are divided into Classical Workloads, using standard CPUs and GPUs, and Quantum Workloads, which employ specialized quantum circuits. This configuration directs tasks to suitable hardware for either real-time edge responses or advanced quantum computations.

Users send tasks to the Orchestrator, which distributes them across a system, balancing energy use and performance. Tasks are routed through Edge-Fog-Cloud to Classical or Quantum hardware based on complexity for sustainability. Simple tasks stay at the Edge to reduce data transfer energy, while complex problems may go to Quantum processors, solving in seconds what would take classical servers weeks.

\section{Energy Efficiency and Green Advantage}
\label{sec:EEGA}

Energy Efficiency has become the standard for defining the "Green Advantage" in computing, especially when quantum elements are added to hybrid quantum classical systems. While classical supercomputers like Frontier use over 21 MW, a quantum-integrated node consumes around 25 kW for cryogenic and control systems. This advantage arises from the reversibility of quantum logic; unlike classical bits that generate heat when erased (Landauer’s Principle) \cite{b18}, quantum algorithms can 'uncompute,' allowing energy costs to scale linearly or sub-linearly with problem complexity. For tasks like catalyst design or grid optimization, this results in an exponential reduction in Joules used, bypassing the thermal "power wall' slowing classical scaling. The table \ref{tab:QEC} shows a operational Energy Comparison to Solve a Problem, that demonstrates a key sustainability metric favoring quantum-enhanced architectures.

\begin{table}[htbp!]
    \centering
        \renewcommand{\arraystretch}{1.5} % Adds padding between rows for readability

    \caption{Operational Energy Comparison to Solve a Problem}
    \label{tab:QEC}
    \begin{tabular}{|p{3.0cm}|p{4.5cm}|p{4.5cm}|}
        \hline
        \scriptsize{\textbf{Metric}} & \scriptsize{\textbf{HPC (Classical)}} & \scriptsize{\textbf{Integrating Quantum-Classical Fabric}} \\ 
        \hline
        \scriptsize{Peak Power} & \scriptsize{20-30 MW} & \scriptsize{25kW(QPU) +50 kW (Edge Node)} \\
        \hline
        \scriptsize{Time Consumption}  & \scriptsize{12 hours (Heuristic Approximation)} & \scriptsize{45 seconds (Near-Optimal)} \\
        \hline
        \scriptsize{Energy Consumption}   & \scriptsize{$\approx 300,000\text{ kWh}$} & \scriptsize{$\approx 0.9\text{ kWh}$} \\ 
        \hline
        \scriptsize{Carbon Footprint} & \scriptsize{$\approx 120\text{ metric tons} CO_2$} & \scriptsize{$\approx 0.0004\text{ metric tons } CO_2$} \\
        \hline
    \end{tabular}
\end{table}

The Green Performance Advantage includes carbon-aware orchestration that uses quantum devices as precision accelerators, as shown in the table \ref{tab:QEC} By offloading only the most demanding tasks, like AI sampling or algorithm optimization, it is possible to decrease carbon footprint considerably. The use of SDN, for example will improve this by routing tasks to quantum nodes powered by local renewables, making the Quantum-Classical Fabric a carbon-neutral resource. This integration ensures high performance supports global decarbonization without sacrificing environmental sustainability \cite{b19}.

\begin{figure}[!htbp]
  \begin{center}
    \includegraphics[width=0.60\textwidth]{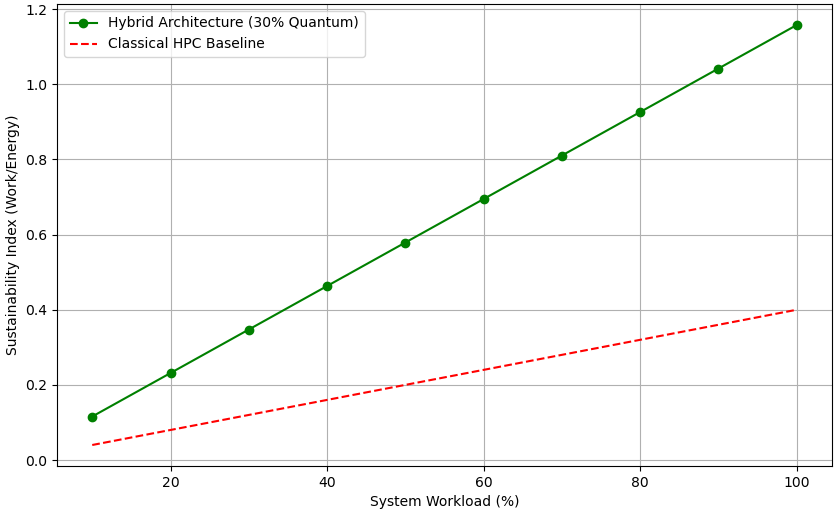}
  \end{center}
  \caption{Sustainability Index vs System Load for Hybrid Quantum-Classical and Classical HPC Architectures}
  \label{fig:SISI2}
\end{figure}

Figure \ref{fig:SISI2} shows that as system workload rises, a hybrid quantum–classical architecture consistently outperforms a purely classical HPC baseline in terms of sustainability index (work per unit energy). The performance gap widens at higher loads. Although both systems benefit from spreading fixed overheads over increased workload, the hybrid approach, in this case with about 30$\%$ quantum offloading, cales more effectively. Quantum resources accelerate the most demanding subproblems, lowering iteration counts and total energy consumption for solutions. Overall, this results in a two- to three-fold boost in sustainability across the full workload spectrum, demonstrating that targeted, energy-conscious integration of quantum computing can offer a significant green performance edge for large, optimization-intensive applications like digital twins.

In other words, a Quantum-Classical Computing hybrid architecture consistently outperforms the classical HPC baseline in terms of sustainability index across all workload levels, as shown in the Figure \ref{fig:SISI2}. As system load increases, the difference becomes more pronounced, demonstrating that the hybrid approach uses energy more efficiently to do meaningful work, especially at medium to high utilization. This trend emphasizes a significant eco-friendly benefit: with higher work output per unit of energy, the hybrid system can lower overall energy use and carbon emissions while improving performance. This makes it a more sustainable choice for future high-performance computing tasks.

\begin{figure}[!htbp]
  \begin{center}per
    \includegraphics[width=0.65\textwidth]{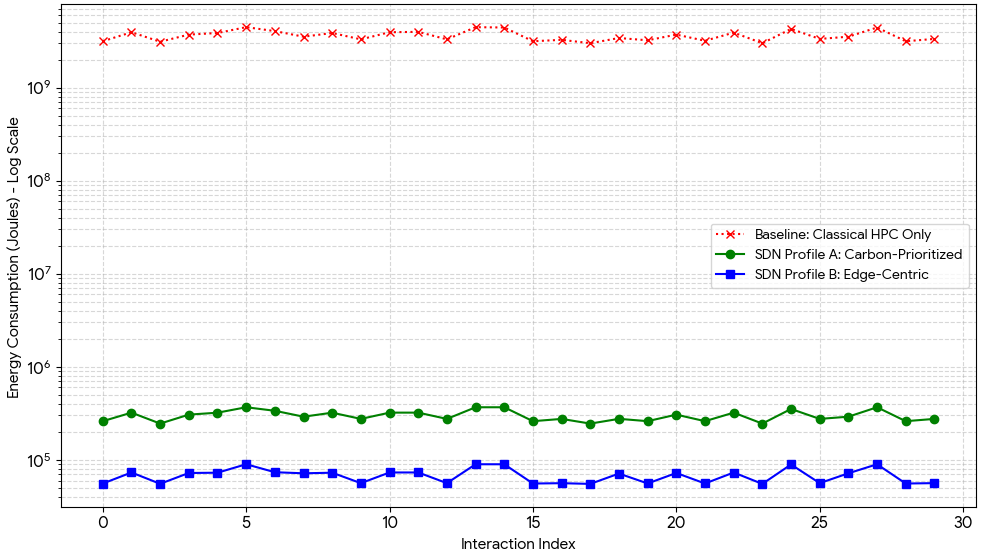}
 \end{center}
 \caption{Comparison of Computing Strategies: HPC-Classical Baseline vs SDN-Profiles}
  \label{fig:SDNP}
  \end{figure}

A comparison of energy use among different computing approaches shows that SDN-enhanced profiles greatly surpass the traditional HPC baseline un the Figure \ref{fig:SDNP}. As shown in the figure, the "Classical HPC Only" baseline has a high energy consumption close to Joules, while the "Carbon-Prioritized" (Profile A) and "Edge-Centric" (Profile B) methods cut energy use by roughly four orders of magnitude, bringing it down near Joules. These findings indicate that using SDN orchestration to focus on carbon savings or edge computing significantly reduces the energy impact of large-scale operations compared to classic centralized systems.

%Figure \ref{fig:SDN} shows an SDN-based, carbon-aware routing that dynamically directs quantum workloads to renewable-powered quantum resources while keeping classical tasks at the edge. Simulations over 30 interactions show this approach significantly reduces energy use and carbon emissions compared with a classical HPC baseline, demonstrating the green benefits of quantum-enabled computing. The framework models 4 quantum edge devices, 12 classical edge devices, a classical HPC center, and a large quantum system. In 30 interactions with 30$\%$ quantum edge workloads, the quantum continuum consistently cut energy consumption by several orders of magnitude compared with a classical-only HPC baseline.

%\begin{figure}[!htbp]
%  \begin{center}per
%    \includegraphics[width=0.58\textwidth]{SDNBasedCAHPCQC.png}
%  \end{center}
%  \caption{Energy advantage of SDN-orchestrated Quantum–Classical Computing}
%  \label{fig:SDN}
%\end{figure}

The results show a systematic energy reduction by several orders of magnitude when quantum devices are integrated into the computing continuum and controlled with carbon-aware SDN policies. The "Classical HPC Only" baseline maintains a high, static energy demand averaging about $3.88 \times 10^9$ J, with minimal variation, remaining the most energy-intensive. Conversely, the "Carbon-Prioritized" (Profile A) and "Edge-Centric" (Profile B) SDN strategies reduce energy use by about four orders of magnitude to $3.24 \times 10^5$ J and $7.43 \times 10^4$ J, respectively. Statistical analysis shows low, stable energy consumption despite workload fluctuations, with a coefficient of variation between 14$\%$ and 15 $\%$. These findings confirm that SDN orchestration focusing on carbon efficiency or edge processing can greatly reduce energy use in large-scale systems without causing operational volatility compared to traditional centralized approaches.

\section{Further Work and Conclusions}
\label{sec:FWC}

Further research should validate the Adaptive Quantum Classical Fusion (AQCF) framework across additional real-world edge-to-cloud scenarios to improve dynamic task decomposition. While various architectural approaches can manage the computing continuum, Software-Defined Networking (SDN) offers distinct advantages for Distributed Quantum Computing (DQC). By treating the network as a unified computational fabric, SDN centralizes control signal and path management, enabling dynamic resource allocation across the cloud-edge continuum. This orchestration supports critical performance features, such as "warm-start" commands that pre-activate cloud QPUs to reduce latency for edge-detected tasks. Furthermore, the SDN controller enables routing policies that prioritize energy efficiency and carbon-aware computing, ensuring a "Green" performance advantage for the entire system. Specifically, it aims to develop advanced Software-Defined Networking (SDN) policies that prioritize "carbon-neutral" routing by assigning high-demand quantum workloads to nodes powered by renewable energy. Additionally, integrating microwave-to-optical transducers and quantum repeaters will be key to deploying distributed modular quantum nodes, ensuring stable remote entanglement and reducing cooling energy costs during network delays.

This proposal presents a sustainable, energy-efficient method for integrating quantum computing into existing classical infrastructure, with a focus on Quantum Edge Devices to improve the computing continuum. It employs a layered hierarchy, from edge sensing to expansive Quantum Processing Units (QPUs) in the cloud, thereby creating a "Green Advantage" despite heterogeneity. This structure enables the solution of complex problems with lower energy consumption and emissions than traditional high-performance computing (HPC). By closely integrating efficient algorithms, such as classical pre-processing, the framework shows that a hybrid quantum-classical system can dramatically enhance performance while minimizing environmental impact.


\begin{thebibliography}{8}


\bibitem{b1} D. Kimovski et al., "Beyond Von Neumann in the Computing Continuum: Architectures, Applications, and Future Directions," in IEEE Internet Computing, vol. 28, no. 3, pp. 6-16, May-June 2024, doi: 10.1109/MIC.2023.3301010.

\bibitem{b1a} N. Furkan Bar, H. Yetíş, N. Özbey and M. Karaköse, "Quantum Computing Based Collaborative Optimum Mission Allocation Approach for Heterogeneous Multi Unmanned Vehicles," in IEEE Access, vol. 12, pp. 169069-169078, 2024, doi: 10.1109/ACCESS.2024.3498345.

\bibitem{b1b} Ullah, M.H., Eskandarpour, R., Zheng, H., Khodaei, A.: Quantum computing for smart grid applications. IET Gener. Transm. Distrib. 16, 4239–4257 (2022). https://doi.org/10.1049/gtd2.12602

\bibitem{b1c} Ahmed A. Abd EL-Latif, Bassem Abd-El-Atty, Eman M. Abou-Nassar, Salvador E. Venegas-Andraca, Controlled alternate quantum walks based privacy preserving healthcare images in Internet of Things, Optics and  Laser Technology, Volume 124, 2020, 105942, ISSN 0030-3992, \url{https://doi.org/10.1016/j.optlastec.2019.105942}.

\bibitem{b1d} A. Gupta, M. Kumar Maurya, K. Dhere and V. Kumar Chaurasiya, "Privacy-Preserving Hybrid Federated Learning Framework for Mental Healthcare Applications: Clustered and Quantum Approaches," in IEEE Access, vol. 12, pp. 145054-145068, 2024, doi: 10.1109/ACCESS.2024.3464240.

\bibitem{b1e} Panyaram S. Intelligent Manufacturing with Quantum Sensors and AI A Path to Smart Industry 5.0. IJETCSIT [Internet]. 2025 May 18 [cited 2026 Feb. 9];:140-7. Available from: \url{https://www.ijetcsit.org/index.php/ijetcsit/article/view/191}.

\bibitem{b1f} Vanessa García Pineda, Alejandro Valencia-Arias, Francisco Eugenio López Giraldo, Edison Andrés Zapata-Ochoa, Integrating artificial intelligence and quantum computing: A systematic literature review of features and applications, International Journal of Cognitive Computing in Engineering, Volume 7, 2026, Pages 26-39, ISSN 2666-3074, \url{https://doi.org/10.1016/j.ijcce.2025.08.002}.

\bibitem{b1g} Martín-Cuevas, R., and Calleja, G. (2025). Hybrid Quantum-Classical Computing Architectures. In Quantum Technology Applications, Impact, and Future Challenges (pp. 97-106). CRC Press.

\bibitem{b2} A. Furutanpey et al., "Architectural Vision for Quantum Computing in the Edge-Cloud Continuum," in 2023 IEEE International Conference on Quantum Software (QSW), Chicago, IL, USA, 2023, pp. 88-103, doi: 10.1109/QSW59989. \url{https://doi.ieeecomputersociety.org/10.1109/QSW59989.2023.00021}.

\bibitem{b3} M. R. Uddin, S. Shaon, R. Rahman, D. C. Nguyen, O. A. Dobre and D. Niyato, "Quantum Federated Learning: A Comprehensive Survey," in IEEE Communications Surveys and Tutorials, vol. 28, pp. 3942-3975, 2026, doi: 10.1109/COMST.2025.3644750.

\bibitem{b4} M. Chehimi, S. Y. -C. Chen, W. Saad, D. Towsley and M. Debbah, "Foundations of Quantum Federated Learning Over Classical and Quantum Networks," in IEEE Network, vol. 38, no. 1, pp. 124-130, Jan. 2024, doi: 10.1109/MNET.2023.3327365.

\bibitem{b5} S. Dustdar, V. C. Pujol and P. K. Donta, "On Distributed Computing Continuum Systems," in IEEE Transactions on Knowledge and Data Engineering, vol. 35, no. 4, pp. 4092-4105, 1 April 2023, doi: 10.1109/TKDE.2022.3142856.


\bibitem{b5b} Pournazari, J., Ullah, A., Al-Dubai, A. et al. Computation offloading in the edge-to-cloud compute continuum: a survey of federated architectural solutions. Cluster Comput 28, 839 (2025). \url{https://doi.org/10.1007/s10586-025-05577-6}.

\bibitem{b6} Danelutto, M., Dazzi, P., Torquati, M. (2024). Structuring the Continuum. In: Barolli, L. (eds) Advanced Information Networking and Applications. AINA 2024. Lecture Notes on Data Engineering and Communications Technologies, vol 203. Springer, Cham. \url{https://doi.org/10.1007/978-3-031-57931-8_21}.

\bibitem{b7} M. Yousif, "Intelligence in the Continuum," 2022 Cloud Continuum, Los Alamitos, CA, USA, 2022, pp. 1-2, doi: 10.1109/CloudContinuum57429.2022.00007. 

\bibitem{b8} V. C. Pujol, P. Raith and S. Dustdar, "Towards a new paradigm for managing computing continuum applications," 2021 IEEE Third International Conference on Cognitive Machine Intelligence (CogMI), Atlanta, GA, USA, 2021, pp. 180-188, doi: 10.1109/CogMI52975.2021.00032. 

\bibitem{b9} Manish Adhikari. (2022). Hybrid Computing Models Integrating Classical and Quantum Systems for Enhanced Computational Power: A Comprehensive Analysis. Journal of Advanced Computing Systems , 2(12), 1-9. \url{https://doi.org/10.69987/}.

\bibitem{b10} M. Dayarathna, Y. Wen and R. Fan, "Data Center Energy Consumption Modeling: A Survey," in IEEE Communications Surveys and Tutorials, vol. 18, no. 1, pp. 732-794, Firstquarter 2016, doi: 10.1109/COMST.2015.2481183.

\bibitem{b11} Manish Adhikari. (2022). Hybrid Computing Models Integrating Classical and Quantum Systems for Enhanced Computational Power: A Comprehensive Analysis. Journal of Advanced Computing Systems , 2(12), 1-9. \url{https://doi.org/10.69987/}.

\bibitem{b12} Phillipson, F., Neumann, N., Wezeman, R. (2023). Classification of Hybrid Quantum-Classical Computing. In: Mikyška, J., de Mulatier, C., Paszynski, M., Krzhizhanovskaya, V.V., Dongarra, J.J., Sloot, P.M. (eds) Computational Science – ICCS 2023. ICCS 2023. Lecture Notes in Computer Science, vol 14077. Springer, Cham. \url{https://doi.org/10.1007/978-3-031-36030-5_2}.

\bibitem{b13} Narges Mehran, Dragi Kimovski, Hermann Hellwagner, Dumitru Roman, Ahmet Soylu, and Radu Prodan. 2024. Scheduling of Distributed Applications on the Computing Continuum: A Survey. In Proceedings of the 16th IEEE/ACM International Conference on Utility and Cloud Computing (UCC '23). Association for Computing Machinery, New York, NY, USA, Article 54, 1–6. \url{https://doi.org/10.1145/3603166.3632540}.

\bibitem{b14} P. Mantha, F. J. Kiwit, N. Saurabh, S. Jha and A. Luckow, "Pilot-Quantum: A Middleware for Quantum-HPC Resource, Workload and Task Management," 2025 IEEE 25th International Symposium on Cluster, Cloud and Internet Computing (CCGrid), Tromsø, Norway, 2025, pp. 01-10, doi: 10.1109/CCGRID64434.2025.00070.

\bibitem{b15} V. De Maio, M. Kanatbekova, F. Zilk, N. Friis, T. Guggemos and I. Brandic, "Training Computer Scientists for the Challenges of Hybrid Quantum-Classical Computing," 2024 IEEE 24th International Symposium on Cluster, Cloud and Internet Computing (CCGrid), Philadelphia, PA, USA, 2024, pp. 626-635, doi: 10.1109/CCGrid59990.2024.00075

\bibitem{b16} Marcello Caleffi, Michele Amoretti, Davide Ferrari, Jessica Illiano, Antonio Manzalini, Angela Sara Cacciapuoti, Distributed quantum computing: A survey, Computer Networks, Volume 254, 2024,
110672, ISSN 1389-1286, \url{https://doi.org/10.1016/j.comnet.2024.110672}.

\bibitem{b17} W. Xia, Y. Wen, C. H. Foh, D. Niyato and H. Xie, "A Survey on Software-Defined Networking," in IEEE Communications Surveys and Tutorials, vol. 17, no. 1, pp. 27-51, Firstquarter 2015, doi: 10.1109/COMST.2014.2330903.

\bibitem{b18} Chattopadhyay, P., Misra, A., Pandit, T., and Paul, G. (2025). Landauer principle and thermodynamics of computation. Reports on Progress in Physics. DOI 10.1088/1361-6633/add6b3

\bibitem{b19} A. Dayf, M. El-Aasser and T. Elshabrawy, "Software-Defined Networking (SDN) Optimizations for Green Multisite Data Centers (GMDC)," 2025 International Conference on Machine Intelligence and Smart Innovation (ICMISI), Alexandria, Egypt, 2025, pp. 89-93, doi: 10.1109/ICMISI65108.2025.11115544.


\end{thebibliography}
%


\end{document}